\documentclass[10pt, aip, jcp, preprint, english, twocolumn, notitlepage]{revtex4-1}
\usepackage[utf8]{inputenc}
\usepackage{amsmath}
\usepackage{amssymb}
\usepackage{bm}
\usepackage{xcolor}
\usepackage{graphicx}
\usepackage{float}
\usepackage{multirow}

\definecolor{blue1}{RGB}{0,0,128}
\usepackage[pdftitle={}
            pdfauthor={},
            pdfsubject={},
            colorlinks, 
            linkcolor=blue1,
            citecolor=blue1, 
            filecolor=blue1,
            urlcolor=blue1]{hyperref}
\usepackage{cleveref}
\crefname{equation}{eq.}{eqs.}%
\crefname{table}{Table}{Tables}%
\crefname{figure}{Fig.}{Figs.}%
\crefname{chapter}{chapter}{chapters}%
\crefname{section}{section}{sections}%
\crefname{appendix}{appendix}{appendices}%
\crefname{enumi}{point}{points}%

\usepackage[inline,draft,silent,nomargin]{fixme}
\fxsetup{theme=color}
\definecolor{fxnote}{RGB}{200,0,0}
\definecolor{fxtarget}{RGB}{30,30,255}

\newcommand{\angstrom}{\text{\normalfont\AA}}
\newcommand{\norm}[1]{\left\lVert#1\right\rVert}

\begin{document}
\title{
GPU-Accelerated Approximate Kernel Method for Quantum Machine Learning}
\date{\today}
\author{Nicholas J. Browning}
\email{nickjbrowning@gmail.com}
\affiliation{Institute of Physical Chemistry and National Center for Computational Design and Discovery of Novel Materials, Department of Chemistry, University of Basel, Klingelbergstrasse 80, CH-4056 Basel, Switzerland}
\author{Felix A. Faber}
\affiliation{Department of Physics, University of Cambridge, United Kingdom}
\author{O. Anatole von Lilienfeld}
\affiliation{Machine Learning Group, Technische Universität Berlin, 10587 Berlin, Germany}
\affiliation{Berlin Institute for the Foundations of Learning and Data, 10587 Berlin, Germany.}

\begin{abstract}
Conventional kernel-based machine learning models for \textit{ab initio} potential energy surfaces, while accurate and convenient in small data regimes, suffer immense computational cost as training set sizes increase. We introduce QML-Lightning, a PyTorch package containing GPU-accelerated approximate kernel models, which reduces the training time by several orders of magnitude, yielding trained models within seconds. QML-Lightning includes a cost-efficient GPU implementation of FCHL19, which together can yield energy and force predictions with competitive accuracy on a microsecond-per-atom timescale. Using modern GPU hardware, we report learning curves of energies and forces as well as timings as numerical evidence  for select legacy benchmarks from atomisitic simulation including QM9, MD-17, and 3BPA.
\end{abstract}

\maketitle

\section{Introduction}
Data-driven approximate machine learning (ML) methods have become increasingly prominent in theoretical chemistry in recent years\cite{Huang2021,vonLilienfeld2020,ceriotti_review_jcp,vonLilienfeld2020_nvc,Ceriotti2021_cr}. In particular, supervised learning can be used to augment accurate, but computationally prohibitive electronic structure calculations. For tasks such as \textit{ab initio} molecular dynamics (AIMD), approximate variantes of the electronic Schr\"{o}dinger equation are solved for every coordinate update. 
Consequently, surrogate ML models which can partially substitute for the quantum calculations are extremely beneficial for reducing the overall computational burden (and carbon foot-print).
In general, these ML models first transform atomic coordinates into an intermediate symmetry-preserving representation, which is then passed on to a non-linear machine learning model, usually one based on neural network (NNs)\cite{behler_hdnn} or kernel methods\cite{rasmussen}. In particular, kernel methods, while accurate and straightforward to train, are typically marred by inferior computational efficiency. This is mainly due to the explicit dependence of the interpolation on every training item in the entire training data set which results in a large number of matrix-vector and matrix-matrix products. The issue becomes particularly cumbersome when gradients or higher order derivatives are to be included in the models' loss function. 
While some efforts have been made to reduce the computational cost of kernel based QML models~\cite{operators_qml, OQML, sGDML, SOAP_opt}, there still remains the formally cubic scaling with the training data set size itself, implying an inherent numerical limitation.

In this work an approximate kernel method, Random Fourier Features (RFF)\cite{RFF} is briefly discussed, and a more computationally efficient variant, termed Structured Orthogonal Random Features (SORF)\cite{SORF}, is introduced. These methods do not rely on all  training points as basis functions when performing inference. Instead, they use a lower-dimensional feature map to approximate shift-invariant kernels which significantly improves the computational cost of evaluation. 

We also provide a software package to perform training and prediction of resulting Quantum Machine Learning (QML) models, termed QML-Lightning. It includes GPU implementations of both RFF and SORF models, as well as a GPU implementation of FCHL19\cite{OQML}, an accurate atom-centered representation. QML-Lightning is built upon the PyTorch software package, with additional CUDA C implementations for critical components to improve its computational throughput. A thorough benchmark of the predictive accuracy of QML-Lightning has been performed on several established datasets of chemical compounds from literature, comparing against existing kernel- and neural-network based models. To assess the models' performance across chemical compound space, we've benchmarked the model against the QM9 dataset\cite{QM9}, which contains 134k small organic molecules with elements C, H, O, N and F. For dynamics and structural relaxation applications, we've benchmarked against both the MD17\cite{MD17_1, MD17_2, MD17_3}, and rectified rMD17\cite{anders_md17} datasets, which contain trajectories of 10 small organic molecules. Finally, to infer the models extrapolative performance QML-Lightning has been benchmarked against the challenging 3PBA dataset\cite{3PBA}, which contains three sets of MD trajectories at temperatures 300K, 600K and 1200K. We note that we have focused on energetic properties in this work, as these are the most critical for AIMD applications, however other QM properties can be used within the QML-Lightning framework straightforwardly. Finally, we provide training times for variety of systems, and prediction times for small molecules as well as periodic systems with up to $\sim$ 17k atoms.

Note that while finalising work on this paper, Dhaliwal \textit{et al.}\cite{Dhaliwal2022} most recently published randomised feature-based interatomic potentials for molecular dynamics with promising results for CPU based applications.

\section{Software Availability}
The QML-Lightning software is provided under an MIT licence at https://github.com/nickjbrowning/QMLightning.

\section{Theory}
\label{sec:theory}
This section first summarises a subset of kernel methods to learn quantum mechanical properties. First, Gaussian Process Regression (GPR)\cite{GAP_tutorial,GPR_for_mats} is introduced to learn both energies and forces of chemical compounds. Then, operator quantum machine learning (OQML)\cite{operators_qml, OQML} is discussed, and finally the approximate kernel methods random Fourier features (RFF)\cite{RFF} and structured orthogonal random features (SORF)\cite{SORF} are introduced.

In the following, upper case indices denote the index of chemical compounds in a dataset, while lower-case indices denote the index of atomic centres in each chemical compound.

\subsection{Gaussian Process Regression (GPR)}
In GPR kernel models of quantum mechanical properties, one constructs a linear model using a basis of kernel functions $k(p_j, \cdot)$. For example, to learn potential energies a suitable functional form would be,
\begin{align}
U^\text{pred}(\left \{ z_K, \bm{r}_K \right \}_{K \in \mathcal{K}}) &= \sum_i^N \alpha_i \sum_{J \in \mathcal{J}_i} \sum_{K \in \mathcal{K}} k(\bm{\rho}_J, \bm{\rho}_K) \\
K_{ij}^\text{GPR} &= \sum_{I \in \mathcal{I}}\sum_{J\in \mathcal{J}} k(\bm{\rho}_I, \bm{\rho}_J)
\end{align}
where $\bm{\rho}_J$ and $\bm{\rho}_K$ are the atomic representations of atom $J$ and $K$. We note that these atomic representations are functions of the set of local atomic charges and coordinates $\{z, \bm{r}\}$, which has been omitted for brevity. The sets $\mathcal{J}$ and $\mathcal{K}$ contain the set of atoms for training and query molecules $j$ and $k$, respectively. The coefficients $\alpha_i$ are obtained from the following regularised minimisation problem,
\begin{equation}
    L = \sum_i^N (U_i^\text{ref} - U_i^\text{pred})^2 + \lambda\sum_i^N \sum_j^N \alpha_i\alpha_j K_{ij} 
\end{equation}
which has the following solution in matrix form,
\begin{equation}
    \bm{\alpha} = (\mathbf{K}^\text{GPR} + \lambda\mathbf{I})^{-1}\mathbf{U}^{ref}.
\end{equation}
where the hyperparameter $\lambda$ is a small number in order to regularise and ensure numerical stability upon kernel inversion\cite{tikhonov}. For learning potential energies and atomic forces simultaneously, one can construct an expression for the potential energy as follows,
\begin{equation}
\label{GPR_FORCES}
    \begin{pmatrix}
    \mathbf{U} \\ \mathbf{F} 
    \end{pmatrix} = 
    \begin{pmatrix}
    \mathbf{K}^\text{GPR}  & -\frac{\partial}{\partial \vec{r}}\mathbf{K}^\text{GPR}  \\
    -\frac{\partial}{\partial \vec{r}^*}\mathbf{K}^\text{GPR}   &\frac{\partial^2}{\partial \vec{r}\partial \vec{r}^*}\mathbf{K}^\text{GPR}
    \end{pmatrix} \bm{\alpha}
\end{equation}
where matrix notation is introduced for simplicity, and $\frac{\partial}{\partial \vec{r}}$ is used as a shorthand to stack the following derivatives,
\begin{equation}
    \frac{\partial K_{ij}^\text{GPR}}{\partial r_J^l} = \sum_{I\in \mathcal{I}} \sum_{J\in \mathcal{J}} \frac{\partial k(\bm{\rho}_I, \bm{\rho}_J)}{\partial r^l_J}
\end{equation}
where $l$ indexes the coordinate components from the from the query atom $J$. The hessian in equation \ref{GPR_FORCES} has the following form,
\begin{equation}
    \frac{\partial^2 K_{ij}^\text{GPR}}{\partial r_I^k \partial r_J^l} =  \sum_{I\in \mathcal{I}} \sum_{J\in \mathcal{J}} \frac{\partial k(\bm{\rho}_I, \bm{\rho}_J)}{\partial r_I^k \partial r_J^l}
\end{equation}
where $k$ indexes the coordinate components from training atom $I$. The dimension of the full GPR kernel is $(3MN + N)\times(3MN + N)$, where $N$ is the number of training molecules, and $M$ is the average number of atoms per molecule in the entire training set. 
In particular, the Hessian term in equation \ref{GPR_FORCES} has a compute time scaling as $O(36N^2M^4)$, which will severely limit applicability with respect to both: large training set sizes as well as large systems.
\subsection{Operator Quantum Machine Learning (OQML)}
To reduce the computational complexity of GPR models, Christensen \textit{et. al.}\cite{operators_qml,OQML} expanded the potential energy in a basis of kernel functions placed on the atomic environments of each atom in the training set,
\begin{align}
U^\text{pred}(\left \{ z_J, r_J\right \}_{J \in \mathcal{J}}) &= \sum_I^N \alpha_I K_{Ij}^\text{OQML}\\
K_{Ij}^\text{OQML}&= \sum_{J \in \mathcal{J}} k(\bm{\rho}_I, \bm{\rho}_J)
\end{align}
where the index $I$ runs over all atoms in the training set. This extends the number of regression coefficients to the number of atoms in the training set, rather than the number of chemical compounds as for GPR models. Atomic forces can be included in the training scheme, resulting in the following equation in matrix form,
\begin{equation}
    \begin{pmatrix}
    \mathbf{U} \\ \mathbf{F} 
    \end{pmatrix} 
    = 
    \begin{pmatrix}
    \mathbf{K}^\text{OQML} \\
    -\frac{\partial}{\partial \vec{r}^*}\mathbf{K}^\text{OQML}
    \end{pmatrix} \bm{\alpha}.
\end{equation}
We note that, unlike GPR models, the basis does not include gradient kernels when training on gradients is required; these derivatives only appear in the loss function as follows,
\begin{equation}
    L(\bm{\alpha}) = \norm{
    \begin{pmatrix}
    \mathbf{U}_\text{ref} \\ \mathbf{F}_\text{ref} 
    \end{pmatrix} 
    -
    \begin{pmatrix}
    \mathbf{K}^\text{OQML} \\
    -\frac{\partial}{\partial \vec{r}^*}\mathbf{K}^\text{OQML}
    \end{pmatrix} 
    {\bm{\alpha}}}^2.
\end{equation}
This loss function is solved directly using a singular-value decomposition (SVD), in which singular values below a threshold $\epsilon_\text{min}$ are ignored in the solution. By contrast to GPR which has $O(36N^2M^4)$ scaling, the heaviest term in the OQML kernel scales as $O(6N^2M^3)$. Note that both these models scale with $N^2$ in the training data, but differ with respect to pre-factor and with respect to scaling with system size.

\subsection{Random Fourier Features (RFF)}
In order to further reduce the explicit dependence of the model on the amount of training data when performing inference, Rahimi \textit{et. al.}\cite{RFF}, introduced a lower dimensional lifting function $z(x)$ to approximate the inner product synonymous with the kernel method,
\begin{equation}
    k(\bm{\rho}_I, \bm{\rho}_J) = \left < {\phi}(\bm{\rho}_I), {\phi}(\bm{\rho}_J) \right > \approx z(\bm{\rho}_I)^T z(\bm{\rho}_J).
\end{equation}
As a consequence of Bochner's theorem the Fourier transform of a shift-invariant kernel $k(\bm{\rho}_I,\bm{\rho}_J) = k(\bm{\rho}_I-\bm{\rho}_J)$ is a proper probability distribution. Consequently one can readily define an explicit feature map which approximates the kernel via Monte Carlo integral estimation,
\begin{widetext}
\begin{align}
    k(\bm{\rho}_I - \bm{\rho}_J) &= \int_{\mathbb{R}^d} p({\bm{w}}) e^{j \bm{w}^T (\bm{\rho}_I - \bm{\rho}_J)} d{\bm{w}} \\ \nonumber
    &\approx \frac{1}{N_F} \sum_{i=1}^{N_F} e^{j \bm{w}_i^T(\bm{\rho}_I - \bm{\rho}_J)} \\ \nonumber
    &\approx  \left[ \frac{1}{\sqrt{N_F}}e^{j\bm{w}_1^T {\bm{\rho}_I}} \dots  \frac{1}{\sqrt{N_F}}e^{j\bm{w}_{N_F}^T \bm{\rho}_I}\right] \left[ \frac{1}{\sqrt{N_F}}e^{j{\bm{w}}_1^T \bm{\rho}_J} \dots  \frac{1}{\sqrt{N_F}}e^{j{\bm{w}}_{N_F}^T \bm{\rho}_J}\right] \\ \nonumber
    &\approx \bm{z}(\bm{\rho}_I)^T\bm{z}(\bm{\rho}_J).
\end{align}
\end{widetext}
where $N_F$ is number of independent vectors $\bm{w}$ drawn from the probability distribution $p(\bm{w})$. For different kernels, the distribution $p(\bm{w})$ takes different forms, however for Gaussian kernels used here, $p(\bm{w})$ is also Gaussian. This formalism readily yields the following low dimensional feature map,
\begin{multline}
    \mathbf{z}(\bm{\rho}_I) = \sqrt{\frac{2}{N_F}} [ \cos(\bm{w}_1^T \bm{\rho}_I + b_1),\\ \dots,\cos(\bm{w}_{N_F}^T \bm{\rho}_J + b_{N_F}) ]^T
\end{multline}
where $b$ is sampled from a uniform distribution on $[0, 2\pi]$. Since potential energies are extensive, one can partition them into atomic energy contributions, and the representation of the atomic environment is passed into this low-dimensional feature mapping,
\begin{align}
    E_i\left(\{q, \mathbf{r}\}\right) &= \sum_{I \in \mathcal{I}} \epsilon_I \\ \nonumber
    &=
    \sum_{I \in \mathcal{I}} \bm{\alpha}^T \bm{z}(\bm{\rho}_I) \\ \nonumber
    &= \mathbf{Z}\bm{\alpha}.
\end{align}
where $\mathbf{Z} \in \mathbb{R}^{N_\text{train}\times N_F}$ is the feature matrix corresponding to $N_\text{train}$ training observations. The $N_F$ weights $\bm{\alpha}$ are the solution to the following regularised normal equation,
\begin{equation}
    \left (\mathbf{Z}^T \mathbf{Z} + \lambda \mathbf{I}\right) \bm{\alpha} = \mathbf{Z}^T \mathbf{E}.
\end{equation}
where the coefficients are obtained first by an LU decomposition of $\left (\mathbf{Z}^T \mathbf{Z} + \lambda \mathbf{I}\right)$. To include forces in the training scheme, the derivatives of the feature vectors $\frac{\partial z_I^l}{\partial r_i^k}$ are computed, where $l$ and $k$ are the feature and coordinate component indexes, respectively, and stored in a derivatives feature matrix $\mathbf{\partial Z} \in \mathbb{R}^{3N_\text{atom}^\text{total} \times N_F}$. The following regularised normal equation is then solved,\begin{equation}
    \left(\left(\mathbf{Z}, \mathbf{\partial \mathbf{Z}}\right)^T \left(\mathbf{Z}, \mathbf{\partial \mathbf{Z}}\right)+ \lambda \mathbf{I}\right) \bm{\alpha} = \left(\mathbf{Z}, \mathbf{\partial \mathbf{Z}}\right)^T \left(\mathbf{E}, \mathbf{F}\right) 
\end{equation}
where the notation $\left(\mathbf{Z}, \mathbf{\partial \mathbf{Z}}\right)$ indicates the concatenation of the feature matrix $\mathbf{Z}$ with the derivative features $\mathbf{\partial Z}$.  The dominant term in constructing the normal equations is the $\partial \mathbf{Z}^T\partial \mathbf{Z}$ matrix product, which scales as $O(3N_\text{train}M N_F^2)$, where $N_\text{train}$ is the number of training molecules and $M$ is the average number of atoms per molecule in the training set. Note that the cost of constructing the normal equations is now linear in $N_\text{train}$ in both energy-only and energy and force learning.

\subsection{Structured Orthogonal Random Features (SORF)}
The above formulation revolves around computing the linear transformation $\mathbf{W}\bm{\rho}_I$. Storing and computing this linear transformation has $O(N_Fd)$ space and time complexity, where $N_F$ is the number of features and $d$ is the size of the atomic representation vector $\bm{\rho}_I$. To reduce this space-time complexity, Yu. \textit{et al}\cite{SORF} introduced structured orthogonal random features (SORF). In this method, the matrix $\mathbf{W}$ is replaced by a special structured matrix consisting of products of random binary diagonal matrices and Walsh-Hadamard matrices. The resulting linear transformation has $O(N_F\log{d})$ time complexity and $O(d)$ or $O(1)$ space complexity, depending on implementation. Briefly, in the case of Gaussian kernel approximation, one can replace the transformation,
\begin{equation}
    \mathbf{W}_\text{RFF} = \frac{1}{\sigma} \mathbf{G}
\end{equation}
where $\mathbf{G} \in{\mathcal{R}^{N_F\times d}}$ is a random Gaussian matrix, with the following transformation,
\begin{equation}
    \mathbf{W}_\text{ORF} = \frac{1}{\sigma} \mathbf{S}\mathbf{Q}
\end{equation}
where $\mathbf{Q}$ is a uniformly distributed random orthogonal matrix (e.g obtained via QR decomposition of $\mathbf{G}$) and $\mathbf{S}$ is a diagonal matrix with entries sampled i.i.d from the $\chi$-distribution with $d$ degrees of freedom. The resulting matrix $\mathbf{SQ}$ is an unbiased estimator of the Gaussian kernel with low variance\cite{SORF}. While this construction still has $O(N_Fd)$ time complexity as well as the additional cost of computing the QR decomposition in a pre-processing step, one can further approximate this transformation as,
\begin{equation}
    \mathbf{W}_\text{ORF} \approx \frac{\sqrt{d}}{\sigma} \mathbf{Q} \approx \frac{\sqrt{d}}{\sigma} \left[\mathbf{H}_i\mathbf{D}_i\right]_{N_\text{transform}}
\end{equation}

where $\mathbf{S}$ has first been replaced by a scalar $\sqrt{d}$ and the random orthogonal matrix $\mathbf{Q}$ has been replaced by a special type of structured matrix. The brackets indicate that this operation is repeated $N_\text{transform}$ times. The matrices $\mathbf{D}_i \in \mathbb{R}^{d \times d}$ are diagonal sign-flipping matrices, where each entry is sampled from a Rademacher distribution, and $\mathbf{H}$ is the Walsh-Hadamard matrix,
\begin{align*} 
H_{2^n} &= 
\begin{pmatrix}
H_{2^{n-1}} & H_{2^{n-1}} \\
H_{2^{n-1}} & -H_{2^{n-1}}
\end{pmatrix} \\
H_1 &= 1
\end{align*}
for $n\geq 1$. Note that when the number of features $N_F > d$, the operation is simply repeated $\frac{N_F}{d}$ times, with the resulting vectors concatenated into a length $N_F$ vector. Crucially, the product $\mathbf{W}_\text{SORF} \bm{\rho}_I$ now has time complexity $O(N_\text{transform}N_F\log{d})$, since multiplication with $\mathbf{H}$ can be efficiently implemented via the fast Hadamard transform using in-place operations in $O(d\log{d})$ time. Finally, since the Walsh-Hadamard matrix is only defined in $\mathbb{R}^{2^n \times 2^n}$, $ \bm{\rho}_I$ must also be projected into $2^n$ dimensions. This is achieved via an SVD decomposition on a subset of the atomic environment representations for each element $e$, concatenated into the matrix $\mathbf{\Tilde{Z}}_e$,
\begin{equation}
    \mathbf{\Tilde{Z}}_e = \mathbf{U}_e\mathbf{S}_e\mathbf{V}_e^T
\end{equation}
and the atomic representations are projected into a lower dimension via the following matrix product,
\begin{equation}
    \bm{\rho}_{I}^\text{proj} = \bm{\rho}_I^T\mathbf{U}_\text{e}^{N_\text{PCA}}
\end{equation}
where only the first $N_\text{PCA}$ columns from the matrix $\mathbf{U}_e$ are used. The subscript $e$ indicates that the matrix $\mathbf{\Tilde{Z}}_e$ is built using only atomic representations of atom type $e$, hence each element has its own projection matrix $\mathbf{U}_e^{N_\text{PCA}}$. Here we've found $N_\text{PCA}=128 \:\text{or}\: 256$ to be sufficient. 

Finally, we note that there are a number of other approximate kernel methods which aim to reduce computational complexity, including other RFF-type approximations\cite{fastfood_features,random_binning_features,rff_survey}, as well as those based on the Nystr\"{o}m method\cite{nystrom_method}, which relies on low-rank structure in the kernel matrix. Here, however, we've opted to use SORF due to its simplicity, computational efficiency and accuracy\cite{rff_survey}.
\subsection{Representation}
In this work we use FCHL19\cite{OQML} as the permutationally and rotationally invariant atomic environment featurisation layer. FCHL19 is an atom-centered representation consisting of two- and three-body elemental bins, similar in construction to the atom-centered symmetry functions (ACSFs) of Behler\cite{ACSF_1, ACSF_2}. The functional form is briefly summarised here. For every unique combination of two elements $\mathcal{X}, \mathcal{Y}$, the representation for each atom $i$ is constructed as follows,
\begin{equation}
    G_i(\left\{Z_j, R_j\right\}_{\mathcal{X}, \mathcal{Y}}) = \left [G^\text{2-body}_\mathcal{X},G^\text{2-body}_\mathcal{Y}, G^\text{3-body}_\mathcal{X, Y} \right] 
\end{equation}
where $\left\{Z, R\right\}^{\mathcal{X},\mathcal{Y}}$ refers to the set of atomic charges and coordinates that have either element $\mathcal{X}$ or $\mathcal{Y}$. The two-body function is given by the following,
\begin{multline}
    G^\text{2-body} = f_\text{cut}(r_{ij})\frac{1}{r_{ij}^{N_2}} \frac{1}{R_s \sigma(r_{ij})\sqrt{2\pi}} \times \\
    \exp{(- \frac{(\ln{R_s} - \mu(r_{ij}))^2}{2\sigma(r_{ij})^2})}
\end{multline}
where $R_s$ are the $n_2$ radial grid centres linearly distributed between 0 and $r_\text{cut}$, and $\mu(r_{ij})$ and $\sigma(r_{ij})$ are the parameters of the log normal distribution,
\begin{equation}
\mu(r_{ij}) = \ln{\left( \frac{r_{ij}}{\sqrt{1 + \frac{\eta_2}{r_{ij}^2}}}\right)}
\end{equation}
\begin{equation}
\sigma(r_{ij})^2 = \ln{\left( 1 + \frac{\eta_2}{r_{ij}^2}\right)}
\end{equation}
where $\eta_2$ is a hyperparameter. The cutoff function to smoothly decay the representation to zero at $r_\text{cut}$ is defined as,
\begin{equation}
    f_\text{cut}(r_{ij}) = \frac{1}{2} \left(\cos{\left( \frac{\pi r_{ij}}{r_\text{cut}}\right) + 1}\right).
\end{equation}
The three-body term $G_\mathcal{X, Y}^\text{3-body}$ is given body the following function,
\begin{multline}
    G^\text{3-body} =\xi_3 G^\text{3-body}_\text{radial} G^\text{3-body}_\text{angular} \times \\ f_\text{cut}(r_{ij})f_\text{cut}(r_{jk})f_\text{cut}(r_{ik}).
\end{multline}
The radial term  $G^\text{3-body}_\text{radial}$ is given by the following expression,
\begin{multline}
G^\text{3-body}_\text{radial} = \sqrt{\frac{\eta_3}{\pi}}\times \\ 
\exp{\left ( -\eta_3 \left (  \frac{1}{2} \left ( r_{ij} + r_{ik}\right ) - R_s\right)^2\right )}
\end{multline}
where $\eta_3$ is a parameter that controls the width of the $n_3$ radial distribution functions, located at $R_s$ grid points. The three-body scaling function $\xi_3$ is the Axilrod-Teller-Muto term\cite{muto, axilrod} with modified exponents\cite{fchl18},
\begin{equation}
    \xi_3 = c_3 \frac{1 + 3 \cos{\theta_{kij}}\cos{\theta_{ijk}}\cos{\theta_{jki}}}{\left(r_{ik}r_{jk}r_{ki}\right)^{N_3}}
\end{equation}
where $\theta_{kij}$ is the angle between atoms $k$, $i$, $j$, with $i$ at the centre, $c_3$ is a weight term and $N_3$ is a three-body scaling factor. Finally, the angular term is given by a Fourier expansion,
\begin{equation}
G^\text{3-body}_\text{angular} = \left [ G_n^\text{cos}, G_n^\text{cos}\right ]
\end{equation}
where the cosine and sine terms are given by,
\begin{align}
    G_n^\text{cos} &= \exp{\left ( - \frac{\left(\zeta n\right)^2}{2}\right )} \big( \cos{\left ( n\theta_{kij}\right )} \nonumber \\ 
    &- \cos{\left ( n\left(\theta_{kij} + \pi\right)\right )} \big) \\
    G_n^\text{sin} &= \exp{\left ( - \frac{\left(\zeta n\right)^2}{2}\right )} \big( \sin{\left ( n\theta_{kij}\right )} \nonumber \\
    &- \cos{\left ( n\left(\theta_{kij} + \pi\right)\right )} \big)
\end{align}
where $\zeta$ is a parameter describing the width of the angular Gaussian function, and $n>0$ is the expansion order. Similarly to previous work, only the two $n=1$ cosine and sine terms are used. 

\subsection{Computational Details}
\subsubsection{Optimisation of Representation Parameters}
The optimal parameters to generate the FCHL19 representation differ here than in the original implementation\cite{OQML}. While the energy + force parameters are the same, albeit with a lower cutoff of $r_\text{cut} = 6.0\angstrom$, we have found improved energy-only parameters. To fit these parameters, we employed a subset of 576 distorted geometries of small molecules with up to 5 atoms of the type CNO, saturated with hydrogen atoms, for which forced and energies have been obtained from DFT calculations\cite{operators_qml, Christensen2018_dataset}. This dataset is identical to that used in the original FCHL19 publication. This dataset is randomly divided into a training set of 384 geometries and a test set of 192 geometries. Models are fitted to the training set, and predictions on the test set are used to minimize the following cost function with respect to the parameters,
\begin{multline}
    \mathcal{L} = 0.01 \sum_i (U_i - U_i^\text{ref})^2 + \\
    \sum_i\frac{1}{n_i}\norm{\bm{F}_i - \bm{F}_i^\text{ref}}^2
\end{multline}
where $U_i$ is the energy of molecule $i$ and $\bm{F}_i$ and $n_i$ are the forces and number of atoms for the same molecule. A greedy Monte Carlo optimisation was used to perform this optimisation, where real-type parameters are optimised by multiplying with a factor random chosen from a normal distribution centred on 1 with a variance of 0.05, and integer-type parameters by randomly adding +1 or -1. The final parameters found to work best are listed in table ~\ref{table:fchl19_params}.

\begin{table}[]
\begin{tabular}{lcc}
\hline
Parameter      & E     & E + F \\ \hline
$n_2$          & 23    & 24    \\
$n_3$          & 22    & 20    \\
$\eta_2$       & 0.27  & 0.32  \\
$\eta_3$       & 5.6   & 2.7   \\
$N_2$          & 2.78  & 1.8   \\
$N_3$          & 2.1   & 0.57  \\
$c_3$          & 60.1  & 13.4  \\
$\zeta$         & $\pi$ & $\pi$ \\
$r_\text{cut}$ & 6.0   & 6.0   \\ \hline
\end{tabular}
\label{table:fchl19_params}
\caption{Optimised representation parameters for FCHL19 for both energy-only (E) and simultaneous energy and force (E + F) learning. }
\end{table}

\subsubsection{GPU Implementation: Representation} 
The FCHL19 representation is constructed by assigning each atom $i$ to each block in a batch. One block is launched for each atom in the system. For each block, a total of 256 threads are used in a 2-dimensional thread grid. The first dimension of this grid contains 16 threads, and enumerates over all two-body interactions with central atom $i$ to construct $G_2$, while second dimension contains 8 threads, which enumerates the third index in the three body interaction $G_3$. For the forwards pass, the reduction of all $G_2^i$ and $G_3^i$ scalar elements is performed global memory. For the backwards pass, however, the $\left [f_x^i,f_y^i, f_z^i \right]$  force components are stored and summed in (local, on-chip) memory therefore significantly increasing throughput. A simple tiled neighbour-list is used to linearise the cost of FCHL19 with respect to increasing number of atoms in the local environments, therefore only atoms within the cutoff radius are considered when constructing the representation.
Once the atomic representation has been constructed, it is projected to a lower dimension of size $N_\text{PCA}$ using a matrix obtained from an SVD of a randomly selected subset of atomic representations from the training set. The size of this lower-dimensional vector is constrained to be a power of 2 for the purposes of the Hadamard transform. Each element within the training database has its own projection matrix and is used to project down each atomic representation separately.

\subsubsection{GPU Implementation: Structured Orthogonal Features}
For the SORF forward pass, each block handles all $\frac{N_F}{d}$ hadamard transforms for a single atom to produce a feature vector of the desired length $N_F$. The Hadamard transform itself operates on the projected FCHL19 representation (dimension $d = 2^n, n \geq 1$), after multiplication with the diagonal sign flipping matrix, using a shared-memory butterfly operation which has O($d\log{d}$) complexity. This operation is performed iteratively $N_\text{transform}$ times. In this work either $d=128$ or $d=256$ is used, and $N_\text{transform}$ is set to 2. For the backwards pass, the gradients are stored and reduced in shared memory.

\section{Results and Discussion}
\label{sec:results}
\begin{figure}[H]
  \label{nfeatures_ntransform_conv}
  \includegraphics[width=1.0\linewidth]{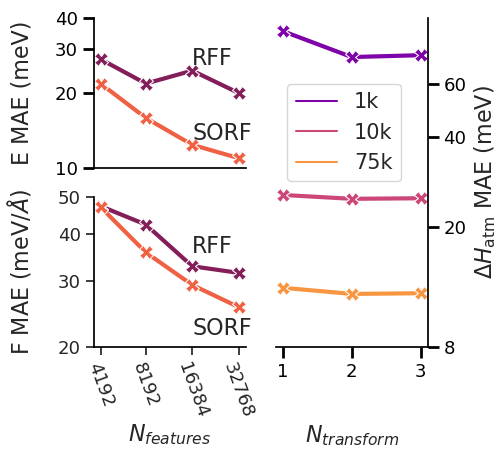}
  \caption
  {Left: Convergence of out-of-sample errors for RFF and SORF models with $N_F$, given a fixed budget of 1000 aspirin configurations, training on both energies and forces. Right: Convergence of out-of-sample errors for SORF models, with $N_\text{transform}=1,2\:\ \text{or} \:3$. The QM9 dataset is used to measure model performance, using 1k, 10k and 75k training samples. Energy units are meV, force units are meV/$\angstrom$.}
\end{figure}
We begin by briefly comparing the performance of RFF-type approximations with those produced from the SORF model used throughout this work. The left column of figure \ref{nfeatures_ntransform_conv} shows the convergence of the out-of-sample energy and force mean absolute errors (MAEs) with respect to increasing $N_F$ used to approximate the kernel. Here, we've used the aspirin trajectory from the unrectified MD17 database\cite{MD17_1, MD17_2, MD17_3}. The amount of training data remains fixed, using $1k$ configurations and training on both energies and forces. While both models display a linear reduction in out-of-sample errors with increasing $N_F$, the SORF model performs notably better for both energies and forces. At 32768 features, the energy and force MAEs for the SORF model are $9$meV and $7$meV/$\angstrom$ lower than that of the RFF model, resulting in a $46\%$ and $19\%$ reduction in relative terms, respectively. This behaviour is consistent across all datasets and systems analysed in this work. We note that in RFF models, increasing $N_F$ incurs significant round-off error in the product $\textbf{W}\textbf{x}$ if performed in FP32 precision, therefore either an error correction scheme or FP64 precision must be used. Conversely, the SORF transform can be performed in FP32 without significant loss in numerical accuracy, culminating in, on average, a two-fold reduction in time required to build the $\textbf{Z}^T\textbf{Z}$ matrix comparatively to RFF models, as well as significant speedup in prediction times on GPUs that do not prioritise FP64 performance. The right column of figure \ref{nfeatures_ntransform_conv} shows the convergence of out-of-sample MAEs on the QM9 database\cite{QM9} of SORF models with the number of transforms, $N_\text{transform}$ used in the SORF featurisation. Here, 1k, 10k and 75k training samples have been used. There is a reasonable improvement of $\approx 11 $meV ($18.3\%$) upon moving from $N_\text{transform}=1$ to $N_\text{transform}=2$ for models trained on 1k samples, however, there is no improvement using $N_{\text{transform}}=3$. For larger datasets, there is a slight improvement for $N_\text{transform} \geq 1$. Based on these findings, $N_\text{transform} = 2$ is used throughout this work. 
\subsection{QM9 Dataset}
\begin{figure}[H]
  \label{QM9}
  \includegraphics[width=1.0\linewidth]{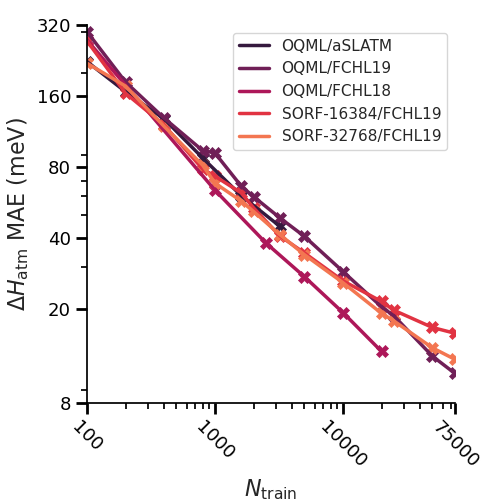}
  \caption{Learning curves for QM9 dataset. The mean absolute error (MAE) of atomisation energy (in meV) is shown for 3 explicit kernel models: OQML with aSLATM, FCHL18 and FCHL19, and two SORF models with FCHL19, using $N_F=16384$ and $32768$, respectively.}
\end{figure}
In figure \ref{QM9} the predictive accuracy of several explicit kernel models and SORF models for atomisation energy of molecules in the QM9 dataset\cite{QM9} are compared. These models include atomic SLATM\cite{SLATM}, FCHL18\cite{fchl18} and FCHL19\cite{OQML}, all using the OQML\cite{OQML} regressor. For the SORF models, learning curves using both $N_F=16384$ and $N_F=32768$ are displayed. We find that the SORF models with FCHL19 perform similarly to OQML with FCHL19: the MAE for OQML/FCHL19 and SORF/FCHL19 at 75000 training samples are $11$meV and $12$meV, respectively. We note that there is a small deviation away from linearity in the learning curve at $\approx50000$ training samples, indicating that more features may be required. This is likely due RFF-type models requiring $N\log{N}$ features\cite{unified_rff} in order to approximate the corresponding kernel.

\subsection{MD-17 and rMD-17 Datasets}
\begin{figure*}
\label{MD-17}
  \includegraphics[width=\textwidth]{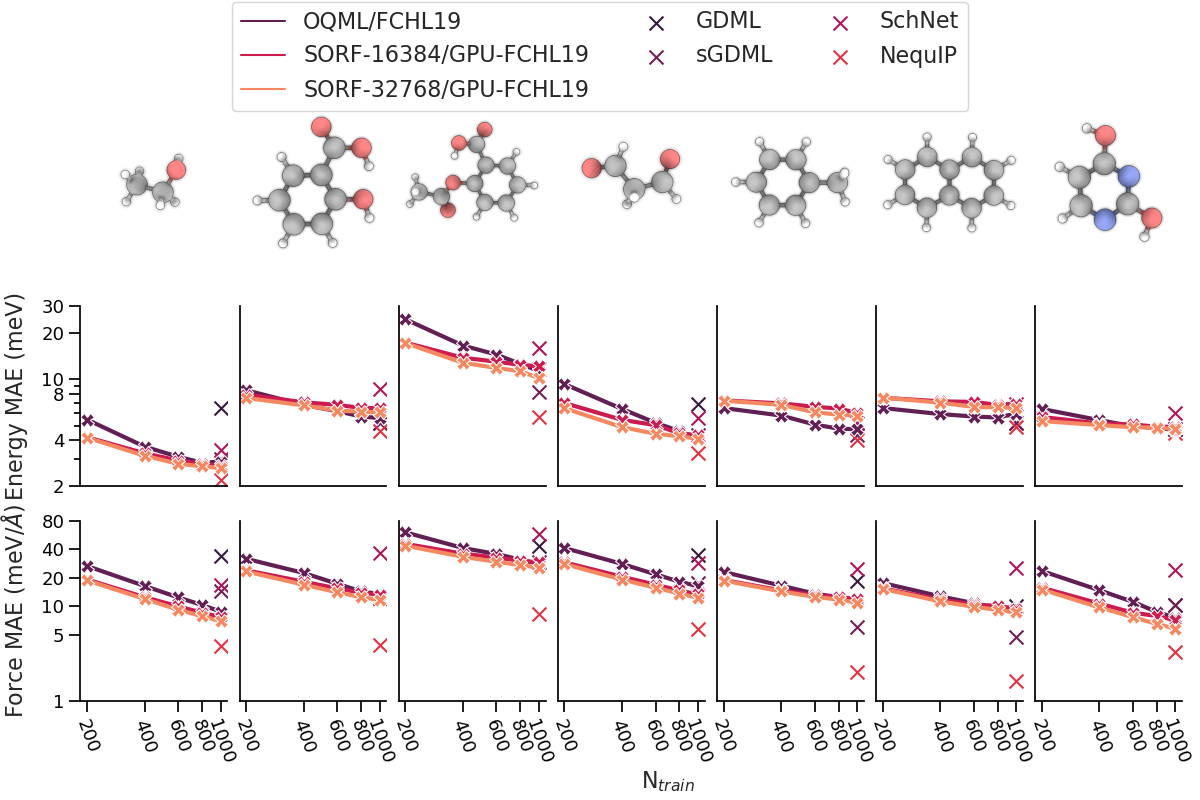}
  \caption{Energy and force learning curves for the molecules (left to right): ethanol, salicylic acid, aspirin, malonaldehyde, toluene, naphthalene and uracil from the MD-17 dataset. The top row contains learning curves for out-of-sample MAE energy prediction (meV). Bottom row contains learning curves for MAE force component prediction (meV/$\angstrom$).}
\end{figure*}
Figure \ref{MD-17} reports the energy and force MAE as a function of number of training samples, using 7 molecules from the MD-17dataset\cite{MD17_1, MD17_2, MD17_3}. To be consistent with previous literature\cite{OQML}, we use the unrectified MD-17 dataset, which is known to contain significant noise on the energy values\cite{anders_md17}. The learning curves for SORF/FCHL19 with both $N_F=16384,\: 32768$ are reported. We compare against OQML models based on FCHL19\cite{OQML}, as well as GDML\cite{GDML} and sGDML\cite{sGDML} models. Additionally, SchNet\cite{schnet} and state-of-the-art NequIP\cite{NEQUIP} neural networks have been included.
\begin{table*}[]
\begin{tabular}{lccccccc}
Molecule                        &        & \multicolumn{1}{c}{SORF-32768} & \multicolumn{1}{c}{GPR/FCHL19} & sGDML & ACE  & \begin{tabular}[c]{@{}c@{}}NequIP\\ (l=0)\end{tabular} & \begin{tabular}[c]{@{}c@{}}NequIP\\ (l=3)\end{tabular} \\ \hline
\multirow{2}{*}{Aspirin}        & Energy & 9.6                             & 6.2                             & 7.2   & 6.1  & 25.2                                                   & \textbf{2.3}                                           \\
                                & Forces & 25.9                            & 20.9                            & 31.8  & 17.9 & 41.9                                                   & \textbf{8.5}                                           \\ \hline
\multirow{2}{*}{Azobenzene}     & Energy & 5.6                             & 2.8                             & 4.3   & 3.6  & 20.3                                                   & \textbf{0.7}                                           \\
                                & Forces & 13.6                            & 10.8                            & 19.2  & 10.9 & 42.3                                                   & \textbf{3.6}                                           \\ \hline
\multirow{2}{*}{Ethanol}        & Energy & 1.5                             & 0.9                             & 2.4   & 1.2  & 2.0                                                    & \textbf{0.4}                                           \\
                                & Forces & 7.5                             & 6.2                             & 16.0  & 7.3  & 13.7                                                   & \textbf{3.4}                                           \\ \hline
\multirow{2}{*}{Malonaldehyde}  & Energy & 2.3                             & 1.5                             & 3.1   & 1.7  & 4.4                                                    & \textbf{0.8}                                           \\
                                & Forces & 12.1                            & 10.3                            & 18.8  & 11.1 & 23.4                                                   & \textbf{5.2}                                           \\ \hline
\multirow{2}{*}{Naphthalene}    & Energy & 4.9                             & 1.2                             & 0.8   & 0.9  & 14.7                                                   & \textbf{0.2}                                           \\
                                & Forces & 9.1                             & 6.5                             & 5.4   & 5.1  & 20.1                                                   & \textbf{1.2}                                           \\ \hline
\multirow{2}{*}{Salicylic acid} & Energy & 4.3                             & 1.8                             & 2.1   & 1.8  & 11.4                                                   & \textbf{0.7}                                           \\
                                & Forces & 12.5                            & 9.5                             & 12.8  & 9.3  & 28.7                                                   & \textbf{4.0}                                           \\ \hline
\multirow{2}{*}{Toluene}        & Energy & 4.2                             & 1.7                             & 1.0   & 1.1  & 9.7                                                    & \textbf{0.3}                                           \\
                                & Forces & 11.6                            & 8.8                             & 6.3   & 6.5  & 27.2                                                   & \textbf{1.6}                                           \\ \hline
\multirow{2}{*}{Uracil}         & Energy & 1.4                             & 0.6                             & 1.4   & 1.1  & 10.0                                                   & \textbf{0.4}                                           \\
                                & Forces & 6.0                             & 4.2                             & 10.4  & 6.6  & 25.8                                                   & \textbf{3.2}                                           \\ \hline
\end{tabular}
\label{rMD-17}
\caption{Energy and force MAEs for models trained on 1k configurations from the revised MD-17 dataset. Errors are reported in meV and meV/$\angstrom$ for energies and forces, respectively.}
\end{table*}

For energy learning, the SORF/FCHL19 models in general display similar accuracies to the OQML/FCHL19 model. For toluene, naphthalene and salicylic acid, the OQML/FCHL19 model slightly outperforms the SORF/FCHL19 models for both $N_F=16384$ and $N_F=32768$. However, in all other cases, the SORF models display similar or better accuracy. both sGDML and GDML perform worse than OQML and SORF models in general, however for toluene and naphthalene specifically, sGDML has the lowest error among the kernel models. 

For force learning, the SORF-based models are as accurate-or-better than OQML/FCHL19 for $N_F=16384$, and reasonably more accurate than OQML for $N_F = 32768$. We note that these models outperform SchNet in all cases, while NequIP out performs the SORF models in all cases. Additionally, as discussed in section \ref{training_times}, we note that the training times times for SORF/FCHL19 are on the order of seconds, while OQML/FCHL19 and sGDML models take several minutes to train. Furthermore, GDML models take several hours, and SchNet and NequiP are trained over hours to days. We additionally provide a comparative benchmark of the revised MD-17 dataset\cite{anders_md17}, a recomputed version of the original MD-17 dataset with tighter SCF convergence criteria. Table \ref{rMD-17} lists the out-of-sample MAEs for the largest SORF model constructed in this work with FCHL19, GPR with FCHL19\cite{anders_md17}, sGDML, ACE\cite{ACE} and NequIP, with rotation orders $l=0$ and $l=3$. We note that FCHL19 is a comparatively simplistic atomic featurisation layer, and consequently it's expected that it does not perform as well as state-of-the-art equivariant many-body neural networks\cite{e3nn, painn, newtonnet} such as NequiP. For a more reasonable comparison the $l=0$ channel NequIP model, which contains at most 3-body terms similarly to FCHL19 has been included here.  We note that while the force errors are similar to the MD-17 results, the energy errors are significantly lower across all models, which is consistent with the observation that the noise floor on the original MD-17 data is higher on the energies.

\subsection{Flexible 3BPA Dataset}
\begin{table*}[t]
\begin{tabular}{lcclccccc}
T (K)                 & Property & SORF-16384 & SORF-32768 & ACE           & sGDML & GAP   & ANI            & ANI-2X \\ \hline
\multirow{2}{*}{300*} & Energy   & 15.0       & 13.7       & \textbf{7.1}  & 9.1   & 22.8  & 23.5           & 38.6   \\
                      & Forces   & 39.2       & 36.2       & \textbf{27.1} & 46.2  & 87.3  & 42.8           & 84.4   \\ \hline
\multirow{2}{*}{600}  & Energy   & 49.3       & 37.7       & \textbf{24.0} & 484.8 & 61.4  & 37.8           & 54.5   \\
                      & Forces   & 86.8       & 75.9       & \textbf{64.3} & 439.2 & 151.9 & 71.7           & 102.8  \\ \hline
\multirow{2}{*}{1200} & Energy   & 118.4      & 99.2       & 85.3          & 774.5 & 166.8 & \textbf{76.8}  & 88.8   \\
                      & Forces   & 175.6      & 159.3      & 187.0         & 711.1 & 305.5 & \textbf{129.6} & 139.6  \\ \hline
\end{tabular}
\label{3BPA}
\caption{Root-mean-squared error of energy and force predictions in meV and meV$/\angstrom$ respectively for different models trained on the 300K 3BPA dataset, using the 600K and 1200K as out-of-distribution test sets.}
\end{table*}
The 3PBA dataset\cite{ACE,3PBA} contains both ambient and high temperature configurational samples for the small drug-like molecule, 3-(bezyloxy)pyridin-2-amine (3BPA). This molecule has 3 central rotatable dihedral angles ($\alpha$, $\beta$, $\gamma$) leading to a complex dihedral potential energy surface with many local minima, which can be challenging for both classical or ML-based potentials\cite{challenges_for_flexible_mols}. In particular, at ambient temperatures the accessible phase space is small, however the dataset contains both 600K and 1200K configurational samples, which have increasingly large phase space volumes. Therefore, a crucial test on whether an ML model can extrapolate well is if the model, when trained on the 300K samples, can accurately predict the 600K and 1200K samples. Consequently, we have trained the SORF/FCHL19 model on the 300K subset of the 3PBA dataset in order to compare against results from ACE, two kernel models sGDML and GAP\cite{GAP_tutorial} using SOAP\cite{SOAP, SOAP_opt}, as well as two related neural network architectures ANI\cite{ANI,torchani} and ANI-2x\cite{ANI-2X}. We note that these results have been summarised directly from the ACE publication\cite{ACE}.

Table \ref{3BPA} lists the energy and force root-mean-squared errors (RMSEs) of a variety of different models listed in the ACE paper\cite{ACE}, including the SORF model with both 16384 and 32768 features. For the 300K test dataset, the SORF model with $N_F=32768$ is able to reach very low errors, having more accurate forces than all models except ACE. For 600K dataset, the SORF models fair less well than both ACE and ANI models, however, they significantly outperform both kernel based models sGDML and GAP with SOAP features. For the 1200K dataset, the SORF model has lower force errors than ACE, however the ANI models have a reasonably lower error. We note that ACE contains up to 5-body terms in its cluster expansion, while FCHL19, which we have not further optimised here beyond reducing the cutoff, contains only up to 3-body terms. Therefore, one would expect ACE to outperform the SORF models in this setting.

\subsection{Timings}
\begin{table*}[t]
\begin{tabular}{lccccccc}
               & \multicolumn{7}{c}{Model Train Times (s)}                                                                                                                                                                                                                                                                             \\ \hline
Device         & 2 x E5-2680                                           & 2 x E5-2680                                          & E5-2640 & V100           & \multicolumn{2}{c}{RTX-3080}                                                                                      & A100                                                    \\ \hline
Molecule       & \begin{tabular}[c]{@{}c@{}}OQML\\ FCHL19\end{tabular} & \begin{tabular}[c]{@{}c@{}}GPR\\ FCHL19\end{tabular} & sGDML   & NeqUIP         & \begin{tabular}[c]{@{}c@{}}SORF\\ 16384-DP\end{tabular} & \begin{tabular}[c]{@{}c@{}}SORF\\ 16384-FP\end{tabular} & \begin{tabular}[c]{@{}c@{}}SORF\\ 32768-DP\end{tabular} \\ \hline
Ethanol        & 66                                                    & 2252                                                 & 144     & $\approx$hours & 24                                                      & 12                                                      & \textbf{4}                                              \\
Salicylic Acid & 249                                                   & 6836                                                 & 282     & $\approx$hours & 46                                                      & 23                                                      & \textbf{7}                                              \\
Aspirin        & 527                                                   & 101451                                               & 570     & $\approx$hours & 84                                                      & 44                                                      & \textbf{10}                                             \\
Malonaldehyde  & 51                                                    & 1926                                                 & 150     & $\approx$hours & 30                                                      & 15                                                      & \textbf{6}                                              \\
Toluene        & 271                                                   & 7976                                                 & 216     & $\approx$hours & 30                                                      & 15                                                      & \textbf{8}                                              \\
Napthalene     & 455                                                   & 11782                                                & 348     & $\approx$hours & 60                                                      & 30                                                      & \textbf{7}                                              \\
Uracil         & 87                                                    & 2576                                                 & 120     & $\approx$hours & 24                                                      & 12                                                      & \textbf{6}                                              \\ \hline
\end{tabular}
\label{training_times}
  \caption{Training time in seconds for various kernel (CPU) and neural network (GPU) models, using $N_\text{train}=1k$ and training on both energies and forces. Device used to train the model indicated where appropriate. For NequIP, training times are approximate. For the QML-Lightning models, both double precision (DP) and single precision (FP) have been specified for the RTX3080 device, whereas only DP has been specified for the A100 (see discussion for details). }
\end{table*}
Table \ref{training_times} shows the total time spent in seconds when training various CPU and GPU-accelerated models using both energies and forces, given a fixed amount of training data  ($N_\text{train}=1k$) from the MD-17 database. The device used to train these models has been listed for reference. Three kernel models are shown, namely OQML and sGDML. We note that OQML and GPR models use the same representation as the GPU variant implemented here. We additionally compare against NequIP, however, the training times shown are an upper bound, as models with good predictive accuracy can likely be obtained with early stopping. The timings for the SORF model with both 16384 and 32768 features are shown, using a consumer RTX3080 GPU and scientific A100 GPU.  It should be noted that there is a significant cumulative round-off error when constructing the $\mathbf{Z}^T\mathbf{Z}$ matrix in floating-point precision (FP32), which leads to ill-conditioning. However, Kahan's summation\cite{kahan} has been implemented here to minimise this error, allowing the normal equations to be constructed in FP32 format to yield a ~3-fold reduction in training time comparatively to FP64 precision. For the A100 GPU, only FP64 performance is shown, as the peak FP64 FLOPs for this card is the same as its FP32 performance, namely due to its support of FP64 to FP64 matrix multiplication via its TensorCore architecture.  Using the RTX3080 GPU, SORF models are several orders of magnitude faster to train than both GPR and neural networks. Furthermore, they are on average $\approx$3 and $\approx$210 times faster to train than OQML and sGDML models respectively across the dataset shown when using FP64 matrix multiplication, and $\approx$10 and $\approx$700 times faster to train when using FP32 with Kahan's summation. On the A100 GPU, remarkably, trained models can be obtained in seconds. We note that this work focuses on models which can fit into GPU global memory, i.e both the $\mathbf{Z}^T\mathbf{Z}$ matrix and the normal equations are constructed and solved on the GPU. However, QML-lightning also supports an out-of-core approach, where the $\mathbf{Z}^T\mathbf{Z}$ matrix is tiled, with each tile being copied to the host device and summed. Finally, the normal equations are then solved on the CPU. This second variant is necessary when the number of features yields matrices which exceed the memory required to both construct the $\mathbf{Z}^T\mathbf{Z}$ matrix, as well as solve the normal equations. In these cases, the choice of CPU(s) or out-of-core CPU/GPU implementation will significantly determine the model training time, however this has not been investigated here.
\begin{table}[t]
\label{timings_mols}
\begin{tabular}{llll}
\hline
               &       & \multicolumn{1}{c}{SORF-16384} & \multicolumn{1}{c}{SORF-32768} \\ \cline{3-4} 
               & Atoms & \multicolumn{2}{c}{Time (per atom) / [ms, ($\mu$s)]}                        \\ \hline
Ethanol        & 9     & 3.2 (355.5)                      & 4.5 (500.0)                    \\
Ethanol (1k) & 9k  & 21.6 (2.4)                     & 39.2 (4.4)                     \\
Aspirin        & 21    & 3.1 (147.6)                      & 4.9 (233.3)                    \\
Aspirin (1k) & 21k & 53.2 (2.5)                     & 86.4 (4.1)                     \\
VUJBEI (s)         & 17k & 85.1 (5.0)                      & 118.3 (7.0)                     \\
H2O (l)         & 12k & 136.4 (11.3)                   & 155.3 (12.9)                     \\ \hline
\end{tabular}
\caption{Prediction time (ms) and cost-per-atom ($\mu$s, parenthesis), for ethanol, aspirin, a periodic metal-organic framework and liquid water. For ethanol and aspirin, both single-configuration and batched configuration performance are listed. All timings listed here were computed on an RTX3080.} 
\end{table}
\begin{figure}[H]
\label{timings_force_call}
  \includegraphics[width=1.0\linewidth]{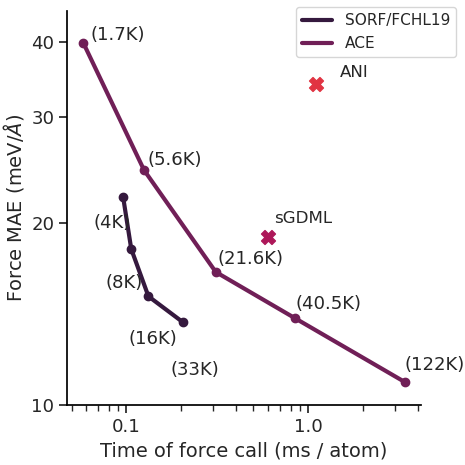}
  \caption{Single-configuration force evaluation time and mean-absolute errors for a variety of models trained on 1K azobenzene configurations on a log-log scale. Parenthesised numbers show the number of features used approximate the kernel for the SORF model, and for ACE they show the number of basis functions. }
\end{figure}
In figure \ref{timings_force_call} we show the force accuracy of the SORF model with FCHL19 as a function of the number of features, and the corresponding evaluation time. For additional context, we compare against ACE, sGDML and ANI models. For the SORF model, an RTX3080 was used, whereas results for ACE\cite{ACE}, sGDML\cite{sGDML} and ANI\cite{ANI} are taken from the ACE paper\cite{ACE}, which used a Xeon Gold 5218. We note that for both the SORF and ANI models, GPU utilisation is low and significantly increases with more configurations. For SORF models in particular, the GPU utilisation also significantly improves with increasing total number of atoms, however here we only show single-configuration performance. Increasing the number of features reasonably improves the error, however, this begins to saturate at 32768 features.

Table \ref{timings_mols} lists the total timings in milliseconds for predicting energies and forces for a variety of systems, with per-atom timings in microseconds in parenthesis. For representative small-molecule timings, both ethanol and aspirin are listed. For larger systems, a brass cluster\cite{brass_1, brass_2}, a metal organic framework (16848 atoms) (refcode: VUJBEI) and liquid water (12000 atoms) are included, where periodic boundary conditions have been implemented using the minimum image convention. For small molecule, single configuration systems, the GPU is significantly underutilised, since each block handles a single atom, there are a significant number of idle streaming multiprocessors (SMs). 
For comparison, timings for simultaneously computing energies and forces for 1000 ethanol and aspirin configurations are presented. Here for the SORF-16384 model, the energy prediction time per configuration reduces from $3.2$ms to $0.022$ms for ethanol, and $3.1$ms and $0.053$ms for aspirin, respectively, clearly showing the effect of increasing atom counts on GPU utilisation. It should be noted that these timings include all CPU and GPU operations, therefore the creation of temporary matrices, host-device and device-host transfers, the device execution time itself as well as CPU and GPU overhead are all contained within. For the MOF system with 16848 atoms, excellent prediction times are obtained, requiring only $85.1$ms and $118.3$ms for $N_F=16848$ and $N_F=32768$, respectively, to compute energies and forces. This results in a cost of $5\mu$s  and $7\mu$s per atom respectively. For liquid water with $12000$ atoms, the total time comparatively increases to $136.4$ms and $155.3$ms for $N_F=16848$ and $N_F=32768$, while the force computation times increase to $11.3\mu$s and $12.9\mu$s per-atom, respectively.
\begin{figure}[H]
  \includegraphics[width=1.0\linewidth]{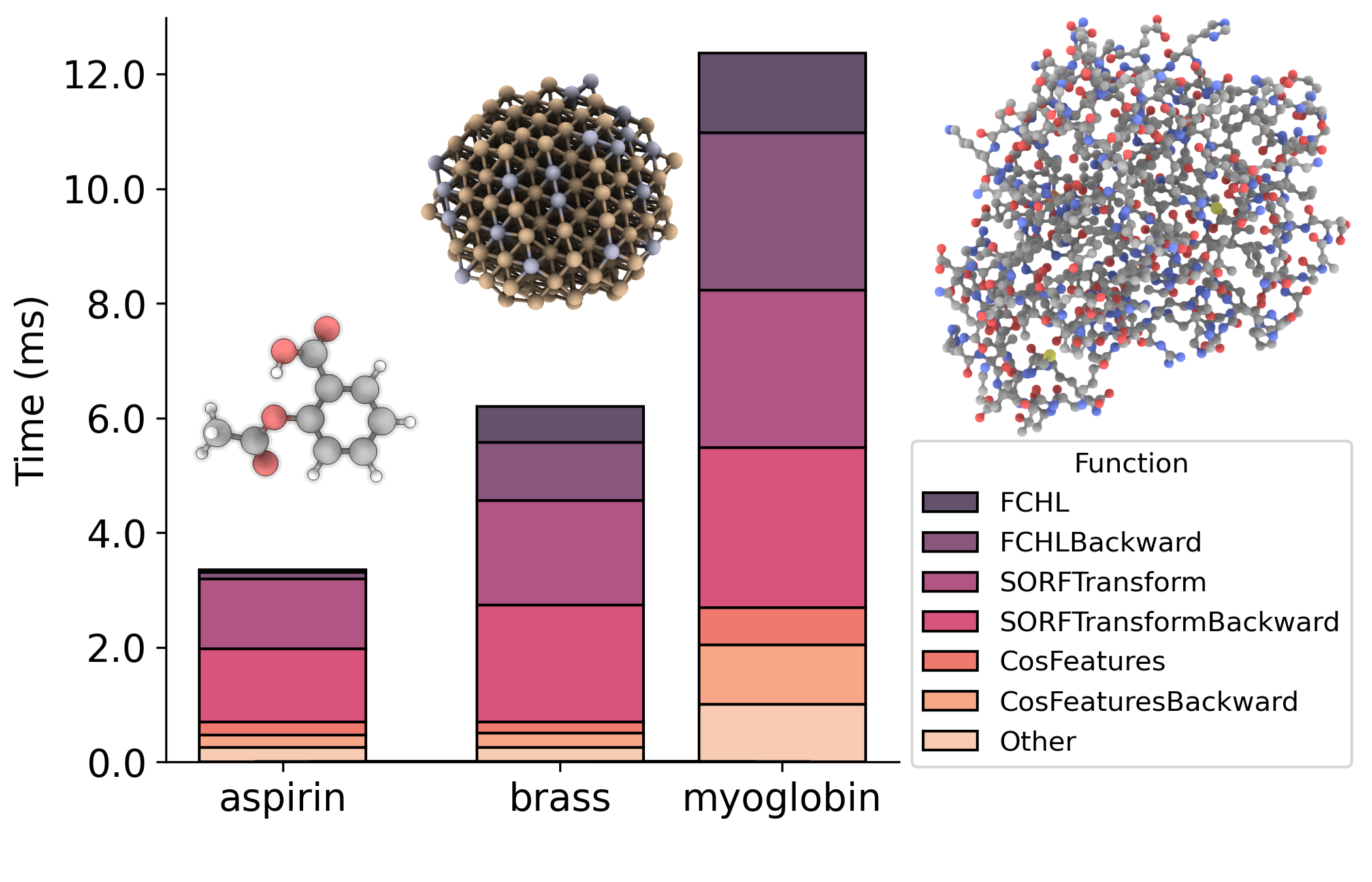}
  \label{component_timing_cost}
  \caption{Average GPU inference (on-device) timings of component functions for energy and force evaluation for A) a single aspirin configuration and B) a single Cu$_{225}$Zn$_{20}$ brass cluster configuration and C) a configuration of myoglobin, with $R_\text{cut} = 6.0\angstrom$ and $N_F = 32768$.}
\end{figure}
For a component breakdown in the computational cost, figure \ref{component_timing_cost} shows the on-device execution time for key components in computing energies and forces for aspirin ($N_\text{atoms} =21$), a brass cluster ($N_\text{atoms}=245$) and myoglobin ($N_\text{atoms} =1260$), using $N_F= 32768$. For aspirin, the dominating costs is the SORF transform, where the forward and backwards pass function cost 1.2ms and 1.3ms, respectively, and together consume ~74.2\% of the total on-device time. The FCHL19 representation costs 0.05ms and 0.1ms for the forward and backwards pass respectively, and the average number of neighbours per-atom is 17. Conversely for the brass cluster, which has over 10-fold the number of atoms as aspirin with an average number of neighbours of 46 (max. 78), the total on-device time only approximately doubles to 6.2ms. Here, the percentage cost of FCHL19 forward and backward passes increases to 26\%, with the SORF transform passes occupying 62.2\% of the total cost. As discussed previously, the sub-linear device time increases is due to poor GPU utilization for small systems. For myoglobin, which has a factor of 5 more atoms than the brass cluster with average number of neighbours of 38 (max 62) this observation is enhanced further, with total device-time approximately doubling again to 12.4ms. In this system, the costs for individual components become more uniform, with the FCHL passes and SORF transforms occupying 33.4\% and 44.7\% of the cost, respectively. 

\section{Conclusion}
In this paper we have introduced a PyTorch-based library, termed QML-Lightning, which contains approximate kernel models and efficient representations designed for learning quantum mechanical properties. We have provided a low-cost, PyTorch-wrapped CUDA C implementation of structured orthogonal random features, a variant of the well-known random Fourier features, as well as a computationally efficient implementation of FCHL19, an accurate atom-centred representation.

The combination of structured orthogonal features and FCHL19 has been benchmarked against existing datasets yielding not only similar-or-better accuracy than explicit kernel models with FCHL19, but also competitive accuracy with contemporary models, with significantly reduced training time and very performant prediction time.

\section{Acknowledgements}
This research was supported by the NCCR MARVEL, a National Centre of Competence in Research, funded by the Swiss National Science Foundation (grant number 182892). Calculations were performed at sciCORE (http://scicore.unibas.ch/) scientific computing center at University of Basel.

\bibliographystyle{ieeetr}
\bibliography{lightning}

\end{document}